\documentclass[11pt]{article}
\usepackage{amsmath,amssymb,amsthm,graphicx}
\newtheorem{Theorem}{Theorem}[section]
\newtheorem{Definition}[Theorem]{Definition}

\newtheorem{Remark}[Theorem]{Remark}

\newtheorem{Example}[Theorem]{Example}
\numberwithin{equation}{section} \allowdisplaybreaks

\renewcommand\abstract{{\bf Abstract}}

\begin{document}
\title{(I,J) similar solutions to Euler and Navier-Stokes equations\footnote{The work is supported by National Natural
Science Foundation of China (No.10861014, No.11161057) and Natural
Science Foundation of Anhui Province university (No.KJ2011Z355,
No.KJ2011Z345) and  Natural Science Foundation of Huainan Normal
University (No.2010LK20). The fax numbers of the corresponding
author: 0086-0871-65123165.}}
\author{Ganshan Yang$^{1,2,+}$}
\date{}
\maketitle
\noindent{\small{${^1}$Department of Mathematics, Yunnan
Nationalities University, Kunming, 650031, China}}\\
{\small ${^2}$ Institute of Mathematics, Yunnan Normal University}\\
{${^+}$ Corresponding author email: ganshanyang@yahoo.com.cn}

\renewcommand{\labelenumi}{[\arabic{enumi}]}

\begin{abstract}
In this paper we introduce (I,J) similar method for incompressible two and three dimensional Euler equations and Navier-Stokes equations, obtain a series of explicit (I,J) similar solutions to the incompressible two dimensional Euler equations, they include all of twin wave solutions, some new singularity solutions, and some global smooth solutions with finite energy. We also discover that twin wave solution and affine solution to two dimensional incompressible Euler equations are respectively plane wave and constant vector. Finally, we supply some explicit piecewise smooth solutions to incompressible three dimensional Euler and an example to incompressible three dimensional Navier-Stokes equations which indicates that viscosity limit of a solution to Navier-Stokes equations does not need to be a solution to Euler equations.
\end{abstract}


{\bf PACS 2010:} 47.10.ad, 47.10.Fg.

\section{Introduction}
In this paper we consider the Euler equations ($\sigma=0$) or Navier-Stokes equations
\begin{equation}\label{1.1}\begin{array}{l}
\left\{
\begin{array}{l}
\frac{\partial u}{\partial t} +(u\cdot\nabla) u +\nabla
p=\sigma\triangle u, \quad \mbox{in}\quad R^n\times(0, T),\\
\mbox{div } u=0,\quad \mbox{in}\quad R^n\times(0, T),~~n=2,~3,
\end{array}
\right.
\end{array}\end{equation}
where $u=u(x,t)=(u_1(x_1,x_2,t),u_2(x_1,x_2,t))$ ans $p=p(x_1,x_2,t)$ denote the velocity and pressure, respectively. Though there is a large amount of physics and mathematics literature
on the Euler and Navier-Stokes equations many basic questions remain open.

There are various open problems in fluid physics. The
Navier-Stokes equation has been recognized as the basic equation and
the very starting point of all problems in fluid physics
(see \cite{Sundkvist}).  One of the most significant developments related
to the above problem may be the discovery of Lax pairs of
two-dimensional and three-dimensional Euler equations (see \cite{FriedlanderVishik})
\begin{equation}\label{1.2}\begin{array}{l}
\omega_t+[\psi,\omega]=0,~~~~\omega=\psi_{x_1x_1}+\psi_{x_2x_2},
\end{array}\end{equation}
where the velocity $u=(u_1,u_2)$ is determined by the stream
function $\psi$ through
\begin{equation}\label{1.3}\begin{array}{l}
u_1=-\psi_{x_1},~~u_2=\psi_{x_2}.
\end{array}\end{equation}
Whether an exact solution to Euler equations is explicitly given via solving vortex equations (the weak Lax pair) to Euler equations? Since the Lax pair has still only weak meaning, then one cannot by Biot-Savart law get the solutions to Euler equations from those solutions of vortex equations. Thus whether the two-dimensional Euler equations are integrable under some stronger meanings similar to those of three-dimensional Euler equations are still open. In this paper we find a so-called (I,J) similar method which can give some explicit smooth solutions to two-dimensional incompressible Euler(see section 2). As application to the (I,J) similar method, a large amount of explicit twin wave solutions are constructed in section 3.

There are various open problems in mathematics. For example, how to establish the global existence of smooth solutions,
or how to establish the blow-up solution at least when the space dimension equals to three (see \cite{Fefferman}). The study of the incompressible Navier-Stokes equations has a long history. A deeper result on the weak solution was obtained by L. Caffarelli, R. Kohn and L. Nirenberg in \cite{CaffarelliKohnNirenberg}. On the blow-up problem of incompressible Navier-Stokes equations, Tsai in \cite{TsaiTP} proved that the Leray self-similar solutions to (\ref{1.1}) must be zero if they satisfy local energy estimates. So in the section 4 we concern with the method of determining the nonexistence of non constant affine solution to the two-dimensional Euler equations, this can justly due to the (I,J) similar method.

Since it is very hard to solve the Navier-Stokes equations in three dimensional space, we consider the two equations in the half space case. In the section 5, we construct some explicit smooth solutions to incompressible three dimensional Euler and Navier-Stokes equations and an example to three dimensional Navier-Stokes equations which indicates that a solution to Navier-Stokes equations need not tend to a solution to Euler equations in the continuous function space on the half space.

On other open problems of the Euler equations and the Navier-Stokes equations, we propose to refer \cite{GIIDPF,MayR} on the incompressible case.

\section{(I,J) similar method of solving the Euler equations and the Navier-Stokes equations}

\begin{Definition} A piecewise smooth solution to (\ref{1.1}) $u(x,t)$ is called a
(I,J) similar solution, if
\begin{equation}\label{2.1}\begin{array}{l}
u(x,t)=\sum\limits_{i=1}^{I}\alpha_i(t)v_i(\sum\limits_{j=1}^{J}\beta_j(t)
M_j(x)),
\end{array}\end{equation}
where $\alpha_{i}(t)$ and $\beta_j(t)$ are smooth functions on $[0,\infty),$\\
 $M_j(x)=(M_{j1}(x), M_{j2}(x),\cdots, M_{jn}(x))$ is a
$n-$dimensional smooth vector function independent of $t,$ and
$v_i(y_1,y_2,\cdots,y_n)$ is a piecewise smooth vector function from $R^n$ to $R^n.$
\end{Definition}
Here a vector value function $f(t)$ is called piecewise smooth on $[0,\infty)$, if
there exist $0< t_1<t_2<\cdots<t_k<+\infty$ such that $f(t)$ is a
smooth function on $(0,\;t_1)$, $(t_i,\;t_{i+1})$, $i=1, 2,
\cdots, k-1$ and $(t_k,\infty)$, respectively. Similarly we say a vector value function $u(x,t)$ to be called piecewise smooth on $R^n$, if there exist $0<r_1<r_2<\cdots<r_k<+\infty$ such that $u(x,t)$ is a
smooth function on $\{x| 0<|x|<r_1\}$, $\{x|r_i<|x|<r_{i+1}\}$, $i=1, 2,
\cdots, k-1$ and $\{x|r_k<|x|<\infty\}$, respectively.
We rewrite
\begin{equation*}\label{}\begin{array}{rl}
(\sum\limits_{j=1}^{J}\beta_j(t)M_j(x))&=(\sum\limits_{j=1}^{J}\beta_j(t)M_{j1}(x),\sum\limits_{j=1}^{J}\beta_j(t)M_{j2}(x),\cdots,\sum\limits_{j=1}^{J}\beta_j(t)M_{jn}(x))\\
&=:(y_1,y_2,\cdots,y_n).
\end{array}\end{equation*}
Insert (\ref{2.1}) into (\ref{1.1}), we have
\begin{equation}\label{2.2}\begin{array}{l}
\sum\limits_{i=1}^{I}\alpha_{it}
v_i+\sum\limits_{i=1}^{I}\sum\limits_{j=1}^{J}\sum\limits_{k=1}^{n}\beta_{jt}
M_{jk}v_{i{y_k}}+\sum\limits_{i_1,i_2=1}^{I}\sum\limits_{j=1}^{J}\sum\limits_{i_2=1}^{n}\alpha_{i_1}\alpha_{i_2}\beta_j
M_{jkx_k}v_{i_2k}v_{i_1y_{k}}+\nabla p\\
=\sum\limits_{i=1}^{I}\sum\limits_{j_1,j_2=1}^{J}\sum\limits_{k,l,m,s=1}^{n}\alpha_i\beta_{j_1}
\beta_{j_2} M_{j_1sx_k}M_{j_2sx_k}v_{iy_k y_l}+\sum\limits_{i=1}^{I}\sum\limits_{j=1}^{J}\sum\limits_{k,l=1}^{n}\alpha_i\beta_jv_{iy_l}M_{jlx_kx_k},\\
\sum\limits_{i=1}^{I}\sum\limits_{j=1}^{J}\sum\limits_{k,l=1}^{n}\alpha_i\beta_jv_{iy_l}M_{jlx_k}=0.
\end{array}\end{equation}

For the incompressible Euler equations, we take $n=2,$ $I=1,$ $J=2,$
$\alpha_1=c(t),$ $\beta_1=\beta_2=1,$ $v_1(y_1,y_2)=(y_1,y_2),$ and
we also set
\begin{equation}\label{2.4}
\begin{array}{c}
M_1(x)=(\frac{x_2}{r^2},-\frac{x_1}{r^2}),~~
M_2(x)=(h(r)x_2,-h(r)x_1),
\end{array}
\end{equation}
where $r=\sqrt{x_1^2+x_2^2}$. Then $u$ and $p$ must satisfy the
following equation:

\begin{equation}\label{2.11}\begin{array}{l}
\frac{c'(t)}{r^2}(x_2,-x_1)-(\frac{c(t)}{r^2}+h(r))^2(x_1,x_2)+\nabla
{p}=0.
\end{array}\end{equation}

So the incompressible Euler equations (\ref{2.2}) have a family of
(I,J) similar solutions
\begin{equation}\label{2.12}\begin{array}{l}
u=((\frac{c(t)}{r^2}+h(r))x_2, -(\frac{c(t)}{r^2}+h(r))x_1),\\
p=-c'(t)\arctan\frac{x_1}{x_2}+F(r,t), x_2\neq 0.
\end{array}\end{equation}
where $F(r,t)=\int r(\frac{c(t)}{r^2}+h(r))^2dr$, $c$ is an
arbitrary smooth function of $t,$ $h$ is an arbitrary smooth function of $r.$

\begin{Theorem} (\ref{2.12}) is a family of (I,J) similar solutions to the incompressible Euler equations
\begin{equation}\label{2.2}\begin{array}{l}
\left\{
\begin{array}{l}
\frac{\partial u}{\partial t} +(u\cdot\nabla) u +\nabla
p=0, \quad \mbox{in}\quad R^2\times(0, T),\\
\mbox{div } u=0,\quad \mbox{in}\quad R^2\times(0, T).
\end{array}
\right.
\end{array}\end{equation}
\end{Theorem}

\begin{Remark}
 To our knowledge, there is little exact solutions to vortex equations,  but they are not solutions to twodimensional Euler equations (\ref{2.2}) except for zero solution and they didn't bring any solution to two dimensional Euler equations (\ref{2.2}) by Biot-Savart law as they singularity (see \cite{TurYanovskyKulik,MajdaBertozzi}), by VIM (see
 \cite{ShangDeng}), and by B\"acklund transformation (see \cite{ShangYadong}). Notice that (\ref{2.12}) is just right a family of exact solutions to two dimensional Euler equations (\ref{2.2}).
\end{Remark}

\begin{Remark}
It is interesting to get many properties by choosing $c(t)$, $h(r)$.
\end{Remark}

\begin{Example}\label{Remark2.4}
According to \cite{MajdaBertozzi},
\begin{equation}\label{}\begin{array}{l}
u=((\frac{c(t)}{|x-Ct|^2}+h(|x-Ct|))x_2+c_1, -(\frac{c(t)}{|x-Ct|^2}+h(|x-Ct|))x_1+c_2),\\
p=-c'(t)\arctan\frac{x_1-c_1t}{x_2-c_2t}+F(|x-Ct|,t),\\
F(r,t)=\int r(\frac{c(t)}{r^2}+h(r))^2dr
\end{array}\end{equation}
is also a solution pair for any constant vectors $C\in R^2$.
$w=Q^Tu(Qx,t), \bar{p}=p(Qx,t)$ is also a solution pair for any
rotation matrixes $Q$.
\end{Example}

\begin{Example}
There are some  $u$ with finite energy only at some points, such as
$t=1.$ Taking $c(t)=t,$ $h(r)=-\frac{1}{r^2}$
\begin{equation}\label{}\begin{array}{l}
u=((\frac{t}{r^2}-\frac{1}{r^2})x_2, -((\frac{t}{r^2}-\frac{1}{r^2})x_1),\\
p=-\arctan\frac{x_1}{x_2}-\frac{1}{2r^2}(t-1)^2.
\end{array}\end{equation}
Then $u(x,0)\in L^2(R^2\backslash B_\delta )\cap
L^{2+\epsilon}(R^2\backslash B_\delta ),$ however for every $\epsilon>0$ and
$\delta>0,$  we have $u(x,t)\in
L^{2+\epsilon}(R^2\backslash B_\delta )\backslash
L^2(R^2\backslash B_\delta ),$ where $B_{\delta}=\{x\in R^2||x|<\delta\}.$
\end{Example}

\begin{Example}
There are some explicit solutions $u$ with singularity only at some points.
Taking $c(t)=\frac{1}{T-t},$ $h(r)=-\frac{1}{r^2}$
\begin{equation}\label{}\begin{array}{l}
u=((\frac{1}{r^2(T-t)}-\frac{1}{r^2})x_2, -(\frac{1}{r^2(T-t)}-\frac{1}{r^2})x_1),\\
p=-\frac{1}{(T-t)^2}\arctan\frac{x_1}{x_2}-\frac{1}{2r^2}(\frac{1}{T-t}-1)^2.
\end{array}\end{equation}
Then $u$ is singular at $r=0$ and blow-up at $t=T.$
\end{Example}

\section{More examples}
In this section we give more explicit nonzero
solutions by considering explicit twin wave solutions to
two-dimensional Euler equations. Here a twin wave solution has a
form of $u=u(x_1-c_1t, x_2-c_2t).$ The twin wave solution is a
(I,J) similar solution. In fact, if we take $I=1,$ $J=3,$
$\beta_1=1,$ $\beta_2=-c_1t,$ $\beta_3=-c_2t,$ $M_1=(x_1,x_2),$
$M_2=(1,0),$ $M_3=(0,1),$ $v_1(y_1,y_2)=u(y_1,y_2)$ then
$u=u(x_1-c_1t, x_2-c_2t).$ Insert it into (\ref{2.2}), we have the following theorem.

\begin{Theorem}
If the pressure is independent of $x$, all twin wave solutions to
two-dimensional Euler equations $u=u(x_1-c_1t, x_2-c_2t)$ will be
given by $u(x,t)=(v(c_3x_1-x_2-(c_3c_1-c_2)t)+c_1,
c_3v(c_3x_1-x_2-(c_3c_1-c_2)t)+c_2)$, where $v$ is any functions of
$c_3x_1-x_2-(c_3c_1-c_2)t,$ and $c_1, c_2, c_3$ are arbitrary
constants.
\end{Theorem}

\begin{Example}
By taking $c(t)=1,$ $h(r)=-\frac{1}{r^2}+\frac{1}{(1+r^2)^2}$ in the Example \ref{Remark2.4}, we have
\begin{equation}\label{Example3.2}\begin{array}{l}
u=(\frac{1}{(1+|x-Ct|^2)^2}x_2+c_1, -\frac{1}{(1+|x-Ct|^2)^2}x_1+c_2),\\
p=-\frac{1}{6(1+|x-Ct|^2)^3}.
\end{array}\end{equation}
These are global smooth twin wave solutions pair for any constant vectors $C\in R^2$.  $w=Q^Tu(Qx,t), \bar{p}=p(Qx,t)$ are
also twin wave solutions pair for any rotation matrixes $Q$.
\end{Example}
\begin{Remark}
Above these solutions in (\ref{Example3.2}) are are symmetry only at some domains. In particular,
if $c_1=c_2$, they are symmetry solution for all $t\geq0$, and if $c_1\neq c_2$ they are not symmetry for all $t\neq0$ and symmetry only at $t=0.$ These examples show that the difference between the velocity of flow and its wave speed $u-C$ has a finite energy over $R^2$, i.e. $u-C\in L^2(R^2).$
\end{Remark}

\begin{Example}
There are some form symmetry solutions $u$ only at some domains, for
example, taking $v(\xi)=\frac{1}{|\xi+T(c_1-c_2)|^2},$ $c_3=1,$ then
\begin{equation}\label{3.10}\left\{\begin{array}{l}
u_1=\frac{1}{|x_1-x_2+(c_1-c_2)(T-t)|^2}+c_1,\\
u_2=\frac{1}{|x_1-x_2+(c_1-c_2)(T-t)|^2}+c_2,\\
p=p(t)
\end{array} \right.\end{equation}
are some twin wave solutions to (\ref{2.2}), and form symmetry only at $t=T,$ or static.
\end{Example}

\begin{Example}\label{Example3.3}
Take $v(\xi)=\frac{1}{(1+|\xi|^2)^2}-c_1,$ $c_3=1.$ Then
\begin{equation}\label{}\left\{\begin{array}{l}
u_1=\frac{1}{(1+|x_1-x_2-(c_1-c_2)t|^2)^2},\\
u_2=\frac{1}{(1+|x_1-x_2-(c_1-c_2)t|^2)^2}+c_2-c_1,\\
p=p(t)
\end{array} \right.\end{equation}
are global smooth twin wave solutions to Euler equations (\ref{2.2}) with finite energy on any bounded domain, but $u-C$ with infinite energy over $R^2$ except for static case. If the components of wave speed is equal, then the system is static.
\end{Example}

\begin{Example}\label{Example3.4}
If we take $v(\xi)=\frac{1}{|\xi|^2},$ $c_3=1.$ Then
\begin{equation}\label{}\left\{\begin{array}{l}
u_1=\frac{1}{|x_1-x_2-(c_1-c_2)t|^2}+c_1,\\
u_2=\frac{1}{|x_1-x_2-(c_1-c_2)t|^2}+c_2,\\
p=p(t)
\end{array} \right.\end{equation}
are some twin wave solutions with singularity to Euler equations (\ref{2.2}).
\end{Example}

\begin{Remark}
Above these solutions in (\ref{Example3.3}) have singular on the line $\{(x_1, x_2)|x_1-x_2=(c_1-c_2)t\}$ for every $t\geq 0$. In particular, we have the following result.

For every given time $t\ge 0$ and arbitrary  line $\{(x_1,
x_2)|Ax_1-Bx_2=C, {A^2+B^2}\ne 0\},$ there exist some solutions with
singularity over the line $\{(x_1, x_2)|Ax_1-Bx_2=C, {A^2+B^2}\ne
0\}.$
\end{Remark}

\begin{Example}\label{3.5}
According to \cite{MajdaBertozzi}, $w=u(x-Ct,t)+C$,
$\bar{p}=p(x-Ct,t)$ is also a solution pair for any constant vectors
$C\in R^2$. $w=Q^Tu(Qx,t), \bar{p}=p(Qx,t)$ is also a solution pair
for any rotation matrixes $Q$.
$w=\frac{\lambda}{\tau}(\frac{x}{\lambda},\frac{t}{\tau})$,
$\bar{p}=\frac{\lambda^2}{\tau^2}(\frac{x}{\lambda},\frac{t}{\tau})$
is also a solution pair.
\end{Example}

\section{Nonexistence}

In this section we consider the explicit affine solution to
two-dimensional Euler equations. Here a solution $u(x,t)$ is
called affine solution, if the $u(x,t)$ is denoted as
$u(x,t)=(v_1(\frac{x_1-c_1t }{x_2-c_2
t}),v_2(\frac{x_1-c_1t}{x_2-c_2t})),$ $c_2\ne 0.$ The affine
solution is a (I,J) similar solution. In fact, this is the
case: $\beta_1=1,$ $\beta_2=-c_1t,$ $\beta_3=1,$ $\beta_4=-c_2t,$
$M_1=(x_1,0),$ $M_2=(1,0),$ $M_3=(0,x_2),$ $M_4=(0,1),$
$z_1=\beta_1M_1+\beta_2M_2,$ $z_2=\beta_3M_3+\beta_4M_4,$
$w(z_1,z_2)=v(\frac{z_1}{z_2}).$ We discovery the following theorem.

\begin{Theorem}
All affine solutions must be twin wave solutions. Affine solutions to two-dimensional Euler equations are constant vectors. This is that there don't exist non constant affine solution to the two-dimensional Euler equations
\end{Theorem}

\section{Blow up solutions to three dimensional cases}
Let $B_M=\{x\in R^3||x|< M\}.$ A solution to (\ref{1.1}) is called as smooth blow-up solution
at finite time $T$, if $u_i,\ p\in C^{\infty}([0,T_1]\times B_M)\cap
W^{m_1,\ q_1}(0,T_1; W^{m_2,\ q_2}( B_M))$ for any $T_1\in (0, T),$
any nonnegative integer numbers $\ m_1,\  m_2,$ and any positive
real numbers $q_1$, $q_2$, but
\begin{equation}\label{6.6}
\lim_{t\to T^{-}}\|u\|_{W^{m,\ q}(B_M)}=+\infty,\quad \lim_{t\to
T^{-}}\|p\|_{W^{m,\ q}(B_M)}=+\infty.
\end{equation}
for some nonnegative integer numbers $m$ and positive real numbers
$q$, $M$.

\begin{Example}\label{Example5.1}
Let f(t) be piecewise smooth, then Euler equations (\ref{1.1}) has a
class of solutions
\begin{equation}\label{2.16}
\left\{
\begin{array}{l} u=f(t)(c_{11}x_1+c_{21}x_2+c_{31}x_3,
c_{12}x_1+c_{22}x_2+c_{32}x_3, c_{13}x_1+c_{23}x_2+c_{33}x_3),\\
p=-\frac{1}{2}f^2(t)\sum\limits_{i,\,j,\,k=1}^3 c_{i k}c_{k j}x_i
x_j-\frac{1}{2}f'(t)\sum\limits_{i,j=1}^3c_{ij}x_i x_j
\end{array}
\right.
\end{equation}
for any constants $c_{ij}$ satisfying $c_{ij}=c_{ji},$
$c_{33}=-c_{11}-c_{22}.$ Moreover the sufficient and necessary
condition of $u$ being smooth is that $f$ is smooth. The blow-up
time of $u$ just is the blow-up time of $f$.
\end{Example}

\begin{Example}\label{Example6.1}
Let $0<T \le \infty$, consider the initial-boundary problem for Navier-Stokes
equations
\begin{equation}\label{6.1}\begin{array}{l}
\left\{
\begin{array}{l}
\frac{\partial u}{\partial t} +(u\cdot\nabla) u+\nabla
p=\sigma\triangle u, \quad \mbox{in}\quad R^3\times(0, T),\\
\mbox{div } u=0,\quad \mbox{in}\quad R^3\times(0, T),\\
u(x,0)=u_0=(u_{10},u_{20},u_{30}),\quad p(x,0)=p_0,\\
u(x,t)|_{s=0}=\frac{1}{\sqrt{T-t}}(0,0,3),\\
u_{i0},\; p_0\in C^{\infty}([0,T_1]\times H^+)\\
\mbox{for any } 0<T_1<T,
\end{array}
\right.
\end{array}\end{equation}
where $H^+=\{x\in R^3|s\ge 0\}.$  Suppose
\begin{equation}\label{6.2}\begin{array}{l}
\begin{array}{l}
u_{10}=\frac{1}{\sqrt{T}}(-1+\exp(\frac{1}{12\sigma T}
s^2-\frac{1}{\sigma \sqrt{T}}s)),\\
u_{20}=\frac{1}{\sqrt{T}}(-1+\exp(\frac{1}{12\sigma T}
s^2-\frac{1}{\sigma \sqrt{T}}s)),\\
u_{30}=-\frac{1}{\sqrt{T}}(1+2\exp(\frac{1}{12\sigma T}
s^2-\frac{1}{\sigma \sqrt{T}}s)),\\
p_0=-\frac{1}{T}(\frac{1}{2\sqrt{T}}s+c),\\
s=\sum\limits_{i=1}^3(x_i-x_{i0}).
\end{array}
\end{array}\end{equation}
Then according to \cite{GuoYangPu}, for the arbitrary constant $c,$
(\ref{2.1}) has a class of smooth blow-up solutions at finite time
$T$
\begin{equation}\label{6.3}
u=(u_1,u_2,u_3),\; p,
\end{equation}
where $u_{i},\; p\in C^{\infty}([0,T_1]\times H^+)$ for any
$0<T_1<T.$
\begin{equation}\label{6.4}\begin{array}{l}
\left\{
\begin{array}{l}
u_1=\frac{1}{\sqrt{T-t}}(-1+\exp(\frac{1}{12\sigma(T-t)}
s^2-\frac{1}{\sigma \sqrt{T-t}}s)),\\
u_2=\frac{1}{\sqrt{T-t}}(-1+\exp(\frac{1}{12\sigma(T-t)}
s^2-\frac{1}{\sigma \sqrt{T-t}}s)),\\
u_3=-\frac{1}{\sqrt{T-t}}(1+2\exp(\frac{1}{12\sigma(T-t)}
s^2-\frac{1}{\sigma \sqrt{T-t}}s)),\\
p=\frac{1}{T-t}(\frac{1}{2\sqrt{T-t}}s+c).
\end{array}
\right.
\end{array}\end{equation}
Moreover the initial function satisfies the second equation
\begin{equation}\label{6.5}
\mbox{div } u_0=0,\quad \mbox{in}\quad R^3.
\end{equation}
\end{Example}

\begin{Remark}
We observe that the fluid velocity remained unchanged on the boundary $s=0,$ and is independent of the size of the fluid viscosity coefficient $\sigma.$ However, this example indicates two facts: 1. the $C^\infty$ solution to Navier-Stokes
equations need not tend to a solution to Euler equations; 2. $u$ will blows up at finite time $T.$
\end{Remark}

{\bf Acknowledgements} Author is grateful to Professor Boling Guo and Professor Zhouping Xin for their support.

\end{document}